\documentclass[twocolumn,showpacs,pra]{revtex4}
\usepackage{dcolumn}
\usepackage{graphicx}

\begin{document}

\title{Generation of coherent terahertz pulses in Ruby at room temperature}

\author{Elena Kuznetsova$^{1}$, Yuri Rostovtsev$^{1}$,
Nikolai G. Kalugin$^{1}$}
\author{Roman Kolesov$^{1}$, Olga Kocharovskaya$^{1}$,
Marlan O. Scully$^{1,2}$}
\affiliation{$^{1}$Institute for Quantum Studies and Department of Physics, Texas A\&M University, College Station, TX, 77843\\
$^{2}$Applied Physics and Materials Science Group, Eng. Quad., Princeton University, NJ, 08544}

\date{\today}

\begin{abstract}
We have shown that a coherently driven solid state medium can potentially
produce strong controllable short pulses of THz radiation.
The high efficiency of the technique is based on excitation of maximal THz
coherence by applying resonant optical pulses to the medium.
The excited coherence in the medium is connected to macroscopic polarization
coupled to THz radiation. We have performed detailed simulations by solving
the coupled density matrix and Maxwell equations. By using a simple $V$-type
energy scheme for ruby, we have demonstrated that the energy of generated THz
pulses ranges from hundreds of pico-Joules to nano-Joules at
room temperature and micro-Joules at liquid helium temperature, with
pulse durations from picoseconds to tens of nanoseconds.
We have also suggested a coherent ruby source that lases on
two optical wavelengths and simultaneously generates THz radiation. We discussed also
possibilities of extension of the technique to different solid-state materials.

\end{abstract}

\pacs{Gy.42.50}

\maketitle

\section{INTRODUCTION}

The search for efficient, high-power, inexpensive, compact,
and room-temperature methods of generation of
coherent teraherz (THz) radiation is one of the main topics in modern
optoelectronics and photonics~\cite{THz1}.
Its importance is based on the fact that
THz radiation has unique potential for a wide range of applications from
diagnostics of different materials (including semiconductors, chemical
compounds, biomolecules, and biotissues),
imaging (for medical and security purposes), to
remote atmospheric sensing and monitoring,
astronomy, etc.~\cite{THz2,THz3}.

In previous works we have focussed on a new approach to the problem
of generation of short coherent THz pulsed radiation
by taking advantage of dramatic enhancement of the nonlinear response of a
medium via maximal quantum coherence \cite{Zibrov,Yuri-Kolya1,Yuri-Kolya2},
created in atomic and
molecular gases with different level configurations (double-$\Lambda$,
V-$\Lambda$, double-V schemes) by coherent laser radiaiton.

As is well-known, quantum coherence can dramatically change the optical
properties of media.
For example, lasing without population inversion (LWI)~\cite{lwi} and
electromagnetically induced transparency (EIT)~\cite{EIT} have been
demonstrated \cite{LWI-exp,EIT-exp} in coherently driven media. These unusual properties are
employed for generation of electromagnetic radiation of different frequencies
ranging from IR~\cite{Harris-opt-param,Boyd-Scully} to UV~\cite{Harris}, and
gamma-rays~\cite{kocharovskaya99prl}.
Enhancement of coherent Raman scattering via maximal coherence has been
demonstrated experimentally~\cite{sautenkov03pra}, and it has applications to
enhanced real time spectroscopy (FAST CARS)~\cite{MaxCoh}.
Previously, the double $\Lambda$ scheme with near-maximal Raman coherence was
used for highly efficient conversion of blue to ultraviolet
light in $Pb$ vapor \cite{Harris}.
A new type of gas-phase optical parametric oscillator \cite{Harris-opt-param}
was suggested for frequency downconversion
to generate 1.88 $\mu$m radiation in Pb vapor, and efficient
infrared upconversion \cite{Boyd-Scully} was suggested to
convert infrared light with a 100 $\mu$m wavelength to the visible.
The first experimental demonstration supporting these ideas was performed in
\cite{Zibrov} where 5 $\mu$m IR radiaiton was generated in $Rb$ atomic vapor.

Many results in this area have been obtained in gases, but, recently, EIT
has been generalized and extended to solids with a long-lived spin coherence
\cite{kuznetsova02pra}, a class of solid materials, namely, rare-earth and transition metal ion
doped dielectrics
have been suggested as very attractive from the point of view of realization
and applications of EIT.
Several applications of EIT to improve performance of solid state
lasers have been already proposed \cite{kuznetsova04pra,kolesov05pra}.
LWI, first realized in a gas medium \cite{LWI-exp}, recently has been
demonstrated in solids as well \cite{lwi06nature}.

\begin{figure}
\center{
\includegraphics[width=7.5cm]{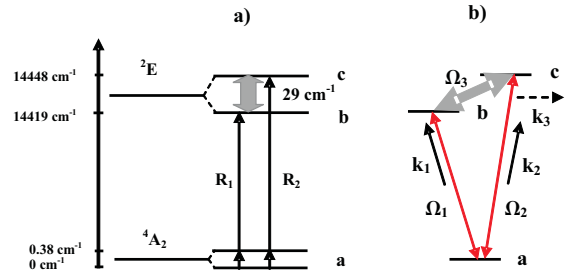}
\caption{\label{fig:ruby-levels}a) Three-level $V$ energy system in ruby proposed for generation of
29 cm$^{-1}$ THz pulses; b) Model $V$ system of energy levels with two co-propagating fields 1 and 2 inducing coherence
between levels $b$ and $c$.}
}
\end{figure}

In this paper we extend the approach of generation of THz radiation via
resonantly induced coherence to solid state media.
The coherence at the corresponding transition can be induced by coherent
optical fields as shown in Fig.~\ref{fig:ruby-levels}.
Electronic levels of the THz transition are coupled by a pair of optical
fields with Rabi frequencies $\Omega_{1}$ and
$\Omega_{2}$ to a common ground state forming a V system of energy
levels. Coherence $\rho_{cb}$ induced by
the optical fields at the THz transition gives rise to polarization provided
that the corresponding dipole
moment is not zero, which will radiate out a THz pulse with Rabi frequency
$\Omega_{3}$. This method differs from
the one proposed in \cite{Yuri-Kolya1,Yuri-Kolya2} for gaseous media where
the optically prepared THz coherence produces no macroscopic polarization
of gaseous media at a THz transition (this is the manifestation of symmetry with
respect to inversion), the reason is that an electrically allowed dipole moment at
the two-photon transition is zero since it is typically forbidden
(although, it is important to note here that the magnetic dipole moment can be non-zero even in
gases, we discuss this opportunity later).
For doped solids having sites with no inversion symmetry is common,
such as the one occupied by Cr$^{3+}$ (simplified energy level structure is
shown in Fig.(\ref{fig:ruby-levels}a)) in ruby,
the corresponding two-photon transition dipole moment is not zero, resulting
in all three transitions in the $V$ scheme being allowed.
It is worth mentioning that using nonlinear optical mixing in a semiconductor
heterostructure for generation of few-cycle THz pulses
was proposed in \cite{Dima-Vitaly}.

In ruby closely spaced R$_{1}$ and R$_{2}$ lines arise from transitions
between the ground state of the
Cr$^{3+}$ ion ($^{4}$A$_{2}$) and its first excited state ($^{2}$E). Each of
these cubic-field states
is split by the trigonal crystal field and spin-orbit
coupling into a pair of
Kramers doublets, 0.38 cm$^{-1}$ apart in the ground state and 29.14 cm$^{-1}$
apart in the $^{2}$E state.
The width of the R lines is about 11 cm$^{-1}$ at room temperature and $0.15$
cm$^{-1}$ at the temperature of liquid nitrogen. The transitions
from the $^{4}$A$_{2}$
ground state to levels $^{2}$E are spin-forbidden, and have oscillator
strength $\sim 7.5\cdot10^{-7}$
(R$_{1}$, R$_{2}$ line peak cross-section is $\sim (1.2-1.4)\cdot 10^{-18}$
cm$^{2}$ at the temperature of liquid nitrogen and
$\sim 4\cdot10^{-20}$ cm$^{2}$ at room temperature).
They are predominantly electric-dipole in nature, since the inversion symmetry
of a cubic crystal field
is broken by crystal field trigonal distortions and odd-parity lattice
vibrations. Namely, the odd-parity component
of the crystal field at the Cr$^{3+}$ ion
mixes odd-parity states of high energy into the even parity d states between
which R-transitions are
observed and makes them weakly allowed. The magnetic-dipole contribution to
the R-lines is estimated to be
$\sim 1/30$ of the strength of the elecric-dipole transitions \cite{ruby}.

The 29 cm$^{-1}$ transition in ruby has history which is as rich as the ruby laser itself.
It was considered promising for realization of a quantum counter
for far-infrared radiation (FIR) \cite{quant-counter} on one hand,
and as a system very convenient to study interaction
processes of non-equilibrium phonons with two-level electronic
systems (so-called phonon spectrometer),
on the other \cite{ruby-phonons}. There was also a proposal to make a
far-infrared laser at this transition pumping via the R$_{2}$ line, but due to
the unfavorable ratio of relaxation rates
(fast relaxation at the FIR and slow at the optical R$_{1}$ transition), the
laser could operate only at liquid helium
temperature and the estimated gain did not exceed losses \cite{FIR-laser}.

It is interesting that 29 cm$^{-1}$ FIR was produced by non-linear mixing of two ruby laser beams, one emitting at
R$_{1}$ and another at R$_{2}$, in LiNbO$_{3}$ \cite{FIR-mixing} and in ZnTe crystals \cite{ZnTe}, with the
efficiency $\eta \sim 10^{-9}$. The method proposed in the present work would allow pulsed THz radiation
to be easily produced from a dual-color ruby laser itself, operating at both R$_{1}$ and R$_{2}$ lines, thus
significantly miniaturizing the system.

The outline of the paper is the following: in the next section we describe a theoretical model used to calculate
parameters of generated THz pulses. In section III we consider possibilities of phase-matching for fields in different
regimes (CW and short pulses). In section IV we make estimates showing the perspective of the suggested scheme as well
as discuss a proposed experiment with a standard femto-second laser setup. Next we show that the system we have considered
is not limited, i.e. there are many systems to which our theory can be applied. Finally we compare the efficiency
of our scheme with the methods of generation of THz radiation used nowadays and possible applications that are opened
due to the high efficiency of the proposed technique.

\section{Theoretical model}
\subsection{Real system}

The ruby system shown in Fig.(\ref{fig:ruby-levels}a) includes eight states, four in the ground $^{4}A_{2}$ and four
in the excited $^{2}E$ one. In our theoretical analysis we describe it using only three levels as in Fig.(\ref{fig:ruby-levels}b),
since this simple model takes into account all essential physics and allows us to make estimates of expected THz radiation
characteristics and required optical field parameters.

\begin{figure}
\center{
\includegraphics{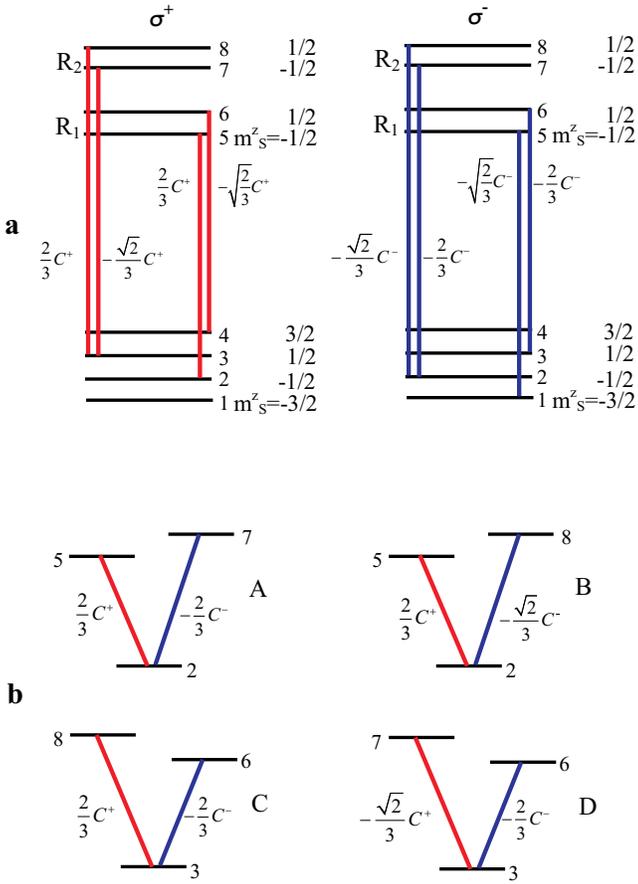}
\caption{\label{fig:sel-rules}a) Selection rules and transition matrix elements for R$_{1}$ and R$_{2}$ optical transitions
in ruby for right and left circular polarizations; b) Four $V$ schemes possible for linearly polarized fields.}
}
\end{figure}

For the proposed technique to work we need to make sure that an efficient $V$ scheme can be realized. In the case of
a single fs pulse, driving both optical transitions simultaneously, we
will have only one polarization, for example, linear. For two pulses we can choose each pulse polarization separately.
Selection rules for the right circular and left circular
polarizations are shown in Fig.(\ref{fig:sel-rules}a) \cite{sel-rules}. As one can see, neither for the right, nor for the
left circular polarization a $V$ scheme involving R$_{1}$ and R$_{2}$ transitions can be organized. For a linear
polarization perpendicular to the optic axis,
which is an equal sum or difference of the right and left ones, there are four $V$ schemes possible, shown in
Fig.(\ref{fig:sel-rules}b).

Taking into account the relation between the coefficients C$^{+}$ and C$^{-}$: $C^{+}C^{-*}=-C^{-}C^{+*}$, one can see
that the schemes A and C cancel each other, since the products of the matrix elements are opposite ($\pm 4C^{+}C^{-*}/9$
for A and C, respectively). On the other hand, for schemes B and D the products are the same ($-2\sqrt{2}C^{+}C^{-*}/9$). These
two $V$ schemes will interfere constructively and induce the THz coherence.

\subsection{Simplified model}

The interaction Hamiltonian for the system shown in Fig.(\ref{fig:ruby-levels}b) is given by
$$
V_{I}=-\hbar \left[\Omega_{1}e^{-i\omega_{ba}t}|b><a|+\Omega_{2}e^{-i\omega_{ca}t}|c><a|+h.c.\right]
$$
\begin{equation}
\label{eq:hamilt}
-\hbar\left[\Omega_{3}e^{-i\omega_{cb}t}|c><b|+h.c.\right],
\end{equation}
where $\Omega_{i}=p_{\mu \nu}{\cal{E}}_{i}/\hbar$ is the Rabi frequency of the respective field, $\mu \nu =ba,\;ca,\;cb$ and
$i=1,2,3$; $p_{ba}$ and $p_{ca}$ are the
electric dipole matrix elements of the optical transitions $b\leftrightarrow a$ and $c\leftrightarrow a$,
and $p_{cb}$ of the THz transition $c\leftrightarrow b$, respectively; $\omega_{ba}$, $\omega_{ca}$ and $\omega_{cb}$ are
the frequencies of the electronic and THz transitions; ${\cal{E}}_{i}$ is the
amplitude of the respective electromagnetic field.

The time-dependent density matrix equations are
\begin{equation}
\label{eq:den-matr}
\frac{\partial \rho}{\partial t}=-\frac{i}{\hbar}[H,\rho]-\frac{1}{2}(\Gamma \rho +\rho \Gamma),
\end{equation}
where $\Gamma$ is the relaxation matrix, and $H$ is the total Hamiltonian.

To estimate the efficiency of THz radiation generation and required laser field intensities we first consider the system (\ref{eq:den-matr}) in the limit
of ultrashort pulses, which are
short compared to relaxation times of the system, and neglect propagation effects for the
optical fields. In this simplified picture
we assume that the system is driven by two resonant fields with equal time-dependent Rabi-frequencies
$\Omega_{1}=\Omega_{2}=\Omega(t)$ (assumed real). In the case of a V-scheme considered in this work the STIRAP
technique, proposed in \cite{Yuri-Kolya1,Yuri-Kolya2} for maximal molecular coherence excitation in a
$\Lambda$ energy system, is not applicable. In the case of the V scheme maximal coherence is excited by two fields
of equal Rabi frequency and duration, acting simultaneouly. The
system is then described by the following set of equations
\begin{equation}
\label{eq:sigca}
\frac{d \sigma_{ca}}{dt}=i\Omega \left(\rho_{a}-\rho_{c}\right)-i\Omega \sigma_{cb},
\end{equation}
\begin{equation}
\label{eq:sigba}
\frac{d\sigma_{ba}}{dt}=i\Omega \left(\rho_{a}-\rho_{b}\right)-i\Omega\sigma^{*}_{cb},
\end{equation}
\begin{equation}
\label{eq:sigcb}
\frac{d\sigma_{cb}}{dt}=i\Omega \left(\sigma^{*}_{ba}-\sigma_{ca}\right),
\end{equation}
\begin{equation}
\label{eq:rhoaa}
\frac{d\rho_{a}}{dt}=i\Omega\left(\sigma_{ba}+\sigma_{ca}\right)-i\Omega\left(\sigma^{*}_{ba}+\sigma^{*}_{ca}\right),
\end{equation}
\begin{equation}
\label{eq:rhobb}
\frac{d\rho_{b}}{dt}=i\Omega\left(\sigma^{*}_{ba}-\sigma_{ba}\right),
\end{equation}
\begin{equation}
\label{eq:rhocc}
\rho_{c}=1-\rho_{a}-\rho_{b}.
\end{equation}
Due to the symmetry of the model system $\rho_{b}=\rho_{c}$, $\sigma_{ba}=\sigma_{ca}$ and $\sigma_{cb}$ - real.
Let us introduce new variables
\begin{equation}
\label{eq:alpha}
\alpha=(\sigma_{ba}+\sigma_{ca})/2=\sigma_{ba},
\end{equation}
\begin{equation}
\label{eq:beta}
\beta=(\sigma_{cb}+\sigma^{*}_{cb})/2=Re(\sigma_{cb})=\sigma_{cb},
\end{equation}
for which, taking into account (\ref{eq:sigcb}) and (\ref{eq:rhoaa}), the following relation can be obtained
\begin{equation}
\label{conservation-law}
\frac{d}{dt}\left(\rho_{a}+2\beta\right)=0,
\end{equation}
giving $\rho_{a}+2\beta=1$.
For the new variables we arrive at the following system of equations
\begin{equation}
\label{eq:alpha1}
\frac{d\alpha}{dt}=i\Omega -4i\Omega \beta,
\end{equation}
\begin{equation}
\label{eq:beta1}
\frac{d\beta}{dt}=2i\Omega\left(\alpha^{*}-\alpha\right),
\end{equation}
which can be further simplified if we substitute $\xi=\alpha-\alpha^{*}$
\begin{equation}
\label{eq:ksi}
\frac{d\xi}{dt}=2i\Omega -8i\Omega \beta,
\end{equation}
\begin{equation}
\label{eq:beta2}
\frac{d\beta}{dt}=-2i\Omega\xi.
\end{equation}
The solution of this system is
\begin{equation}
\label{eq:ksi-sol}
\xi=\frac{\sqrt{2}i}{2}\sin{\left(2\sqrt{2}\int_{-\infty}^{t}\Omega (t')dt'\right)},
\end{equation}
\begin{equation}
\label{eq:beta-sol}
\beta=\sigma_{cb}=\frac{1}{4}\left(1-\cos{\left(2\sqrt{2}\int_{-\infty}^{t}\Omega (t')dt'\right)}\right).
\end{equation}
It follows from Eq.(\ref{eq:beta-sol}) that maximal coherence $\sigma_{cb}=0.5$ will be excited by a pair of
pulses with area $S=2\sqrt{2}\int_{-\infty}^{\infty}\Omega(t)dt=\pi$ each.
Taking pulses of a Gaussian shape $\Omega =\Omega_{0}exp(-t^{2}/2\tau^{2})$ with the area of the pulse
$S=4\sqrt{\pi}\Omega_{0}\tau $, we will have the maximal coherence excited when $\Omega_{0}\tau=\sqrt{\pi}/4\approx 0.443$.
The excited coherence will give rise to the polarization which will radiate out a coherent THz pulse. It is worth
noting that this mechanism is
similar to free-induction decay (FID) \cite{FID} in photon echo: excitation of a maximal coherence in a two-level system by
a $\pi/2$ pulse, followed by emission of a coherent pulse. The difference is that typically in solids the FID decay
is governed by the dephasing due to inhomogeneous broadening of the transition leading to decay of the corresponding
coherence during the time $\sim W_{inh}^{-1}$. 	In the case of ruby the THz transition is homogeneously broadened even
at liquid helium temperature, so the decay of the THz coherence is determined by the homogeneous width of the
corresponding transition.

For two long pulses, which can be considered as CW (in the sense that optical and THz coherences approach
the steady-state), with approximately equal Rabi frequencies $\Omega_{1}\approx \Omega_{2}$,
resonant with the corresponding optical transitions, we get the THz coherence
\begin{equation}
\label{eq:CW-coherence}
\sigma_{cb}=\frac{2\Omega_{2}\Omega^{*}_{1}\exp{\left(-\frac{6|\Omega_{1}|^{2}t}{\gamma +(|\Omega_{1}|^{2}+|\Omega_{2}|^{2})/\gamma_{cb}}\right)}}{\gamma_{cb}\left(\gamma+(|\Omega_{1}|^{2}+|\Omega_{2}|^{2})/\gamma_{cb}\right)},
\end{equation}
where $\gamma_{ba}=\gamma_{ca}=\gamma$ and $\gamma_{cb}$ are the optical and THz coherence decay rates, respectively.
In this case the maximal THz coherence $|\sigma_{cb}|=1$ will be excited provided that $(|\Omega_{1}|^{2}+|\Omega_{2}|^{2})\gg \gamma \gamma_{cb}$,
which is a usual steady-state EIT threshold condition.

Let us now estimate the peak intensity and energy of the optical pulses necessary to excite required coherence.
The peak intensity of the pulse is expressed in terms of the peak Rabi-frequency as
\begin{equation}
\label{eq:intensity}
I_{peak}=\frac{2\pi \hbar c\Omega^{2}_{0}}{\gamma n\lambda \sigma_{abs}},
\end{equation}
where $\gamma$ is the width of an optical transition, $n$ is the refractive index at the optical wavelength,
$\lambda $ is the optical wavelength, and $\sigma_{abs}$ is the corresponding absorption cross-section. The energy
of a single pulse can be calculated as
\begin{equation}
\label{eq:opt-energy}
E_{opt}\approx \frac{2\pi \sqrt{\pi} \hbar c A_{opt}}{\gamma n \lambda \sigma_{abs}}\Omega_{0}^{2}\tau,
\end{equation}
where $A_{opt}$ is the laser beam cross-section.

\section{PROPAGATION: PHASE-MATCHING GEOMETRY FOR FIELDS}

A complete self-consistent calculation also includes field propagation equations
\begin{equation}
\frac{\partial \Omega_{\alpha}}{\partial z}+\frac{n_{\alpha}}{c}\frac{\partial \Omega_{\alpha}}{\partial t}=-\kappa_{\alpha}\Omega_{\alpha}+i\eta_{\alpha}\rho_{\alpha},
\end{equation}
where index $\alpha =1,2,3$ indicates all fields and corresponding polarizations, $\eta_{\alpha}=2\pi \omega_{\alpha}N\mu^{2}_{\alpha}/n_{\alpha}c\hbar$
is the corresponding coupling constant, $\omega_{\alpha}$ are the frequencies of the optical and THz fields, $N$ is the
density of the medium, $c$ is the speed of light in vacuum, $n_{\alpha}$ is the corresponding refractive index,
and $\kappa_{\alpha}$ are losses for the field during propagation in the crystal due to scattering, diffraction,
or non-resonant absorption. For optical fields these losses are usually small, but for the THz field in free space
the diffraction losses given by $\kappa_{3}=\lambda_{3}/D^{2}$ should be taken into account. To avoid diffraction
losses the crystal can be placed in a waveguide for THz radiation, then the distribution of the field mode should be
taken into account.

Propagation effects are important for nonlinear interactions such as wave mixing. Let us consider two regimes: CW,
when the spectral widths of pulses is smaller than the splitting between the levels of the THz transition, and
pulsed one, when the spectral width of pulses exceeds the splitting between the levels.

\begin{figure}
\center{
\includegraphics[width=8cm]{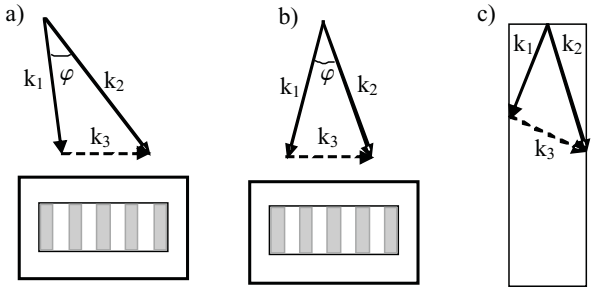}
\caption{\label{fig:phase-matching}a) CW fields 1 and 2 induce coherence between levels $b$ and $c$ in ruby; b)
For short pulses $k_{1}=k_{2}$; c) A three mode ruby laser that generates optical radiation at $R_{1}$ and $R_{2}$ lines
and THz radiation simultaneouly with the ruby crystal serving as a THz waveguide.}
}
\end{figure}

In the CW regime the field $\Omega_{3}$ at the output of the crystal is given by (assuming that the THz pulse
propagates in $z$ direction)
\begin{equation}
\Omega_{3}=ie^{-\kappa_{3}L}\int_{0}^{L}\eta \sigma_{cb}e^{(i\delta k+\kappa_{3})z}dz=i\eta_{3}\sigma_{cb}\frac{e^{i\delta k L}-e^{-\kappa_{3}L}}{i\delta k+\kappa_{3}}=
\end{equation}
$$
=i\eta_{3}\sigma_{cb}Le^{i\delta kL/2}\frac{\sin(\delta kL/2)}{(\delta kL/2)},
$$
if losses are neglected.
Here $L$ is the length of the crystal, and the prefactor describes phase-matching ($\delta k=k_{2z}-k_{1z}-k_{3}$). The
direction of THz propagation is given by the phase-matching condition (see Fig.3)
\begin{equation}
\vec{k_{2}}-\vec{k_{1}}=\vec{k_{3}}.
\end{equation}
To satisfy the phase-matching condition $\delta k=0$ for the THz pulse to propagate in $z$ direction (along the crystal)
the angle between the optical fields has to be adjusted according to the equantion
\begin{equation}
k^{2}_{1}+k^{2}_{2}-2k_{1}k_{2}\cos\phi=k^{2}_{3},\;\cos\phi=\frac{k^{2}_{1}+k^{2}_{2}-k^{2}_{3}}{2k_{1}k_{2}}.
\end{equation}
Rewriting a condition for optical and THz frequencies $\omega_{2}-\omega_{1}=\omega_{3}$ in the form of
$k_{3}=(k_{2}-k_{1})n_{3}/n_{1}$, we obtain
\begin{equation}
\phi=\sqrt{\frac{n^{2}_{3}-n^{2}_{1}}{n^{2}_{1}}}\frac{\lambda_{1}}{\lambda_{3}},
\end{equation}
where $\lambda_{1(3)}$ is the wavelength of the optical (THz) field. For the case of ruby the refractive index for
THz radiation is twice larger than the one for optics, $n_{3}\approx 2n_{1,2}$ \cite{FIR-laser}, then the angle
between directions of propagation for optical fields $\phi=3\cdot 10^{-3}$. The angle between the $k_{1(2)}$
vectors and the THz field propagation direction is $\pi /3$ in this case. This provides an interesting opportunity
to design a dual-wavelength ruby laser working simultaneously at $R_{1}$ and $R_{2}$ lines that can also generate THz
radiation, as is shown in Fig.{\ref{fig:phase-matching}c}. The crystal will simultaneously serve as a THz waveguide
in this case.

Now we turn to a short pulse excitation regime. For many femtosecond laser experimental setups it would be an interesting
extension to include an option of THz generation. The phase-matching in this case can be obtained by excitation of
THz coherence with two optical beams with the same frequencies but different directions of propagation. It is instructive
to show how it works in this case. Let us consider the problem in the simplest case of lowest order of perturbation
in coherent pumping fields. The density matrix equations look like
\begin{equation}
\dot{\sigma_{ab}}=-i\Omega_{1}e^{-ik_{x}x+ik_{z}z}-i\Omega_{2}e^{-ik_{x}x-ik_{z}z},
\end{equation}
\begin{equation}
\dot{\sigma_{ca}}=i\Omega_{1}e^{ik_{x}x-ik_{z}z}+i\Omega_{2}e^{ik_{x}x-ik_{z}z},
\end{equation}
\begin{widetext}
\begin{equation}
\dot{\sigma_{cb}}=i\sigma_{ab}\left(\Omega_{1}(t)e^{ik_{x}x-ik_{z}z}+\Omega_{2}(t)e^{ik_{x}x+ik_{z}z}\right)-
i\sigma_{ca}\left(\Omega_{1}(t)e^{-ik_{x}x+ik_{z}z}+\Omega_{2}(t)e^{-ik_{x}x-ik_{z}z}\right).
\end{equation}
\end{widetext}
A solution of this set of equations is given by
\begin{widetext}
\begin{equation}
\sigma_{cb}=2\int_{-\infty}^{t}dt'\int_{-\infty}^{t'}dt''\left(\Omega_{1}(t')\Omega_{1}(t'')+\Omega_{2}(t')\Omega_{2}(t'')+
\Omega_{1}(t')\Omega_{2}(t'')\cos2k_{z}z+\Omega_{1}(t'')\Omega_{2}(t')\cos2k_{z}z\right).
\end{equation}
\end{widetext}
One can see that the phase-matching between the optical fields and the THz field is achieved by appropriate
tilting of the optical beams at the angle $\phi\approx 2k_{z}/k_{x}=k_{3}/k_{1}=\lambda_{1}n_{3}/\lambda_{3}n_{1}$;
 in particularly, for ruby $\phi\approx 10^{-3}$.

\section{ESTIMATION OF ENERGY OF GENERATED THZ FIELD}

First we analize the optimal conditions at which the maximal THz coherence can be induced in ruby by the optical fields
at both room and liquid helium temperatures.
For room
temperature ruby $\gamma=\gamma_{cb}=5.5$ cm$^{-1}$ \cite{FIR-room-temp}, $n=1.76$,
$\lambda =694.3$ nm, and $\sigma_{abs}=4\cdot10^{-20}$ cm$^{2}$
\cite{ruby}. In the case of ultrashort driving pulses (100 fs-1 ps) the shortest THz pulse that can be generated in this system has a duration $\sim \gamma_{cb}^{-1}\sim 1$ ps.
With the optical pulses of the same duration $\tau=1$ ps the Rabi frequency $\Omega_{0}=4.43\cdot10^{11}$
s$^{-1}$ is required to produce the maximal coherence, corresponding to $I_{peak}\sim 10^{12}$ W/cm$^{2}$, which
is still below a damage threshold of ruby for such duration of the pulses \cite{ruby-damage-threshold}, but we make a
conservative
estimate for lower intensities $I_{peak}\sim 100$ GW/cm$^{2}$, certainly below the threshold, corresponding to the
peak Rabi frequency $\Omega_{0}=2\cdot 10^{11}$ s$^{-1}$. For long driving
pulses (10 ps -1 ns) to excite the maximal coherence
$|\sigma_{cb}|=1$ would require $\Omega_{1,2}^{2}\gg \gamma \gamma_{cb}$
resulting in the Rabi frequency
$\Omega_{1,2} \sim 10^{12}$ s$^{-1}$, corresponding to intensities exceeding the ruby damage threshold for
pulses of such duration. For long pulses (1-10 ns) the damage threshold intensity is
in the range $I_{peak}\sim 20-30$ GW/cm$^{2}$ \cite{ruby-damage-threshold1},
which corresponds to the Rabi frequency $\Omega_{1,2} \sim 10^{11}$ s$^{-1}$. This value is used in the estimates
below.

At low temperatures ($1.8-4.2$ K) the THz coherence lifetime is significantly larger, $\gamma_{cb}^{-1}\sim 500$ ps
\cite{FIR-absorp},
which means that with ultrashort pumping pulses (in this case of 10 ps-1 ns duration), the maximal coherence will
be excited with only $\Omega_{0}\sim 10^{10}$ s$^{-1}$, requiring $I_{peak}\sim 100$ MW/cm$^{2}$ pulse intensity.
For long pulses (1-10 ns) the maximal coherence will be excited when $\Omega_{1,2}^{2}\sim \gamma \gamma_{cb}$,
and since at low temperatures $\gamma \sim 0.05$ cm$^{-1}$ \cite{ruby-R-line}, it results in $\Omega_{1,2} \sim 5\cdot 10^{9}$ s$^{-1}$,
requiring $I_{peak}\sim 30$ MW/cm$^{2}$.

The next question is the energy of the generated THz pulse,
which can be estimated from the propagation equation
\begin{equation}
\label{eq:THz}
\frac{\partial \Omega_{3}}{\partial z}+\frac{n_{3}}{c}\frac{\partial \Omega_{3}}{\partial t}=\frac{2\pi \omega_{3}iN\mu_{cb}^{2}}{n_{3}c\hbar}\sigma_{cb}-\kappa\Omega_{3},
\end{equation}
where $N$ is the density of $Cr^{3+}$ ions, $\hbar \omega_{3}=29$ cm$^{-1}$ ($0.87$ THz), the dipole moment of the far-infrared
transition can be calculated using measured experimentally low-temperature (LT) parameters $\mu_{cb}^{2}=\gamma_{cb}^{LT}c\hbar n_{3}\sigma_{abs,THz}^{LT}/2\pi \omega_{3}$, and
$\kappa$ is the non-resonant absorption coefficient of the sapphire host at the THz wavelength.

Assuming that the crystal sample is side-pumped homogeneously, so that
the THz coherence $\sigma_{cb}$ does not depend
on the propagation coordinate $z$, and phase-matching conditions are satisfied we
arrive at the equation
\begin{equation}
\label{eq:THz1}
\frac{\partial \Omega_{3}}{\partial z}+\frac{n_{3}}{c}\frac{\partial \Omega_{3}}{\partial t}=i\eta\sigma_{cb}(t)-\kappa\Omega_{3},
\end{equation}
where $\eta=N\gamma_{cb}^{LT}\sigma_{abs,THz}^{LT}$, so that at the end of the crystal the Rabi-frequency of the THz pulse is
\begin{equation}
\label{eq:THz-Rabi-fr}
\Omega_{3}(t,L)=i\eta \int_{z_{0}}^{L}\sigma_{cb}\left(t-\frac{z-z'}{c}n_{3}\right)e^{-\kappa\left(z-z'\right)}dz',
\end{equation}
where $z_{0}=max \left(0,L-tc/n_{3}\right)$.
For the ultrashort pumping pulses we can model $\sigma_{cb}(t)=\sigma^{max}_{cb}\exp{(-\gamma_{cb}t)}$ at $t>0$ and $\sigma_{cb}=0$ at
$t<0$ for an estimate.  Eq.(\ref{eq:THz-Rabi-fr})
gives
\begin{equation}
\label{eq:THz-Rabi-short1}
\Omega_{3}=i\eta \sigma^{max}_{cb}\frac{e^{-\gamma_{cb}t}}{\gamma_{cb}n_{3}/c-\kappa }\left(e^{\left(\gamma_{cb}n_{3}/c-\kappa\right)ct/n_{3}}-1\right)
\end{equation}
for $t<Ln_{3}/c$ and
\begin{equation}
\label{eq:THz-Rabi-short2}
\Omega_{3}=i\eta \sigma^{max}_{cb}\frac{e^{-\gamma_{cb}t}}{\gamma_{cb}n_{3}/c-\kappa }\left(e^{\left(\gamma_{cb}n_{3}/c-\kappa\right)L}-1\right)
\end{equation}
for $t>Ln_{3}/c$.

Eqs.(\ref{eq:THz-Rabi-short1})-(\ref{eq:THz-Rabi-short2}) then allow one to calculate the energy of the THz pulse

\begin{equation}
\label{eq:THz-energy}
E_{THz}=\frac{\hbar \omega_{3}A}{\gamma_{cb}^{LT}n_{3}\sigma_{abs,THz}^{LT}}\int_{0}^{\infty}|\Omega_{3}|^{2}dt=
\end{equation}
\begin{widetext}
$$
=\frac{\hbar \omega_{3} AN^{2}\gamma_{cb}^{LT}\sigma_{abs,THz}^{LT}}{n_{3}}\left(\sigma^{max}_{cb}\right)^{2}\left(\frac{1}{2\kappa \gamma_{cb}\left(\kappa+\gamma_{cb}n_{3}/c\right)}-\frac{e^{-2\kappa L}}{2\kappa \gamma_{cb}\left(\gamma_{cb}n_{3}/c-\kappa\right)}+\frac{e^{-\left(\kappa+\gamma_{cb}n_{3}/c\right)L}}{\gamma_{cb}\left(\left(\gamma_{cb}n_{3}/c\right)^{2}-\kappa^{2}\right)}\right),
$$
\end{widetext}
where $A$ is the crystal cross-section along THz pulse propagation direction.
For two pulses of 1 ps duration with $I_{peak}\sim 100$ GW/cm$^{2}$ the maximal induced coherence is $\sigma_{cb}^{max}\approx 0.21$.
Taking $N=1.6\cdot10^{19}$ cm$^{-3}$ for 0.05$\%$ doped ruby, and known from low-temperature measurements
$\gamma_{cb}^{LT}=2\cdot10^{9}$ s$^{-1}$, $\sigma_{abs,THz}^{LT}=3\cdot10^{-19}$ cm$^{2}$
\cite{FIR-absorp}, and room temperature parameters $\gamma_{cb}=10^{12}$ s$^{-1}$ and $\kappa =0.4-0.5$ cm$^{-1}$
\cite{FIR-laser}, $n_{3}=3.5$ \cite{FIR-laser},
and considering the crystal of 1 cm$\times$1 cm$\times$0.1 cm size with the $A=1$ cm $\times$ 0.1 cm THz emitting
cross-section and $L=1$ cm length, we arrive at $E_{THz}\approx 630$ pJ. The energy can be higher for higher optical
peak intensities, approaching several nJ for $I_{peak}\sim 10^{12}$ W/cm$^{2}$. For low temperatures ($1.8-4.2$ K) the parameters are:
$\gamma_{cb}=2\cdot 10^{9}$ s$^{-1}$, $\sigma_{abs,THz}=3\cdot10^{-19}$ cm$^{2}$, $\kappa=0.01$ cm$^{-1}$
\cite{FIR-laser}, and the maximal coherence $\sigma_{cb}=0.5$ can be easily excited. Due to stronger absorption at
low temperatures ($\sigma_{abs}\sim 10^{-18}$ cm$^{2}$ \cite{ruby}), half of the optical intensity will be absorbed
at $\approx 0.5$ mm for $0.05\%$ Cr$^{3+}$ density, therefore, for this estimate we take a crystal with
dimensions 1 cm$\times$0.1 cm$\times$0.05 cm size with the THz emitting cross-section $A=0.1$ cm$\times$ 0.05 cm and
$L=1$ cm length
which results in $E_{THz}\sim 7.5$ $\mu$J.

Let us also make an estimate for fs pulse duration. Since powerful fs Ti:Sapphire lasers are readily available in many
laboratories nowadays this is of interest. In this case the coherence can be excited by just one pulse of 100 fs duration, since its
spectral width is larger than the THz transition splitting. Although, to satisfy phase-matching conditions, it is
necessary to split it in two beams hiting the crystal at slightly different angles. Keeping the same
peak intensity $I_{peak}\approx 100$ GW/cm$^{2}$ we obtain a smaller pulse area $S=0.14$, giving the coherence
$\sigma_{cb}^{max} \sim 2.5\cdot 10^{-3}$ and resulting in THz pulse energy of $\approx 100$ fJ at room temperature.

For the CW pumping case we have, in analogy with the short pulse pumping,
\begin{widetext}
\begin{equation}
\label{eq:THz-energy-CW}
E_{THz}=\frac{\hbar \omega_{3} AN^{2}\gamma_{cb}^{LT}\sigma_{abs,THz}^{LT}}{n_{3}}\left(\sigma^{max}_{cb}\right)^{2}\left(\frac{1}{2\kappa G\left(\kappa+Gn_{3}/c\right)}-\frac{e^{-2\kappa L}}{2\kappa G\left(Gn_{3}/c-\kappa\right)}+\frac{e^{-\left(\kappa+Gn_{3}/c\right)L}}{G\left(\left(Gn_{3}/c\right)^{2}-\kappa^{2}\right)}\right).
\end{equation}
\end{widetext}
where from Eq.(\ref{eq:CW-coherence}) $\sigma^{max}_{cb}=2\Omega^{*}_{2}\Omega_{1}/\left(\gamma \gamma_{cb}+|\Omega_{1}|^{2}+|\Omega_{2}|^{2}\right)$
and $G=6|\Omega_{1}|^{2}/\left(\gamma+\left(|\Omega_{1}|^{2}+|\Omega_{2}|^{2}\right)/\gamma_{cb}\right)$.
In the limit $\Omega_{1,2}^{2}\ge \gamma \gamma_{cb}$
the maximal coherence is excited $\sigma_{cb}=1$, which would require, though, experimentally unfeasible Rabi frequency
$\Omega_{1,2} \sim 10^{12}$ s$^{-1}$ at room temperature, resulting in $I_{peak}\sim 10^{12}$ W/cm$^{2}$. For such
long pulses (10 ps -1 ns) the damage
threshold intensity for ruby is $I^{peak}\sim 20-30$ GW/cm$^{2}$,
which corresponds to the Rabi frequency $\Omega_{1,2}\sim 10^{11}$ s$^{-1}$. For this Rabi frequency
and
room temperature decay rates $\gamma \approx \gamma_{cb}=10^{12}$ s$^{-1}$, the ratio $\Omega_{1,2}^{2}/\gamma \gamma_{cb}\sim 10^{-2}$
 and the resulting THz coherence $|\sigma_{cb}|\sim 10^{-2}$ for a 10 ps pulse. The THz pulse energy will
be $E_{THz}\sim 300$ pJ. At low temperatures the maximal coherence $\sigma_{cb}=1$ can be excited by long pulses (1-10 ns), which would give
energy of a THz pulse $E_{THz}\sim 8$ $\mu$J.

Finally we estimate the total energy of optical pulses using Eq.(\ref{eq:opt-energy}) and overall conversion efficiency to
THz radiation. For room temperature 1 ps pulses pumping the side of the crystal with $A_{opt}=1$ cm$\times$0.1 cm size,
the energy in one pulse according to Eq.(\ref{eq:opt-energy}) is $E_{opt}\approx 27$ mJ, for two pulses, $54$ mJ, respectively. We have to take into account that
only $(1-\exp{(-\sigma_{abs}NL_{opt})})$ of the incident energy is absorbed, where $L_{opt}$ is the size of the
crystal along which the optical pulses propagate. For room temperature $\sigma_{abs}=4\cdot10^{-20}$ cm$^{2}$,
$N=1.6\cdot 10^{19}$ cm$^{-3}$ and $L_{opt}=1$ cm, the absorbed fraction is $0.47$, so about twice the energy calculated
above is required, resulting in $\sim 110$ mJ total energy.
For a pair of 100 fs pulses the same reasoning leads to the total energy of $\sim 11$ mJ ($2.7$ mJ in one pulse from Eq.(\ref{eq:opt-energy})).
For long pulses the total required energy is $300$ mJ.

The expected values of the energy of THz pulses, the peak amplitude of the THz
field if focused to a spot of $300$ $\mu$m size, required
pumping radiation energy and duration and efficiency of optical to THz energy
conversion
for room temperature ruby are summurized in Table \ref{tab:TableI}.

\begin{center}
\begin{table*}
\caption{\label{tab:TableI}Estimates of the THz pulse energy, peak THz field amplitude, required optical pulse
parameters and THz radiation generation
efficiency in ruby at room temperature}
\begin{ruledtabular}
\begin{tabular}{ccc}
\hline
ultrashort & pulses (100 fs-1 ps)  & long pulses (10 ps-100 ps)\\
\hline
100 fs &  1 ps &  10 ps\\
\hline
$\sigma_{cb}\sim 2.5\cdot 10^{-3}$ & $\sigma_{cb}=0.21$ & $\sigma_{cb}\sim 10^{-2}$\\
$E_{THz}\sim 100\;fJ$ & $E_{THz}\sim 630\;$pJ & $E_{THz}\sim 300\;$pJ\\
${\cal{E}}_{THz}\sim 300\;$V/cm & ${\cal{E}}_{THz}\sim 23\;$kV/cm & ${\cal{E}}_{THz}\sim 5\;$kV/cm\\
\hline
$\Omega_{0}=2\cdot 10^{11}\;$s$^{-1}$ & $\Omega_{0}=2\cdot 10^{11}\;$s$^{-1}$ & $\Omega_{0}\sim 10^{11}\;$s$^{-1}$\\
$I_{peak}=100\;$GW/cm$^{2}$ & $I_{peak}=100\;$GW/cm$^{2}$ & $I_{peak}=20-30\;$GW/cm$^{2}$\\
$E_{opt}=110\;$mJ & $E_{opt}=11\;$mJ & $E_{opt}\sim 300\;$mJ\\
$^{a}\eta \sim 6\cdot 10^{-9}$ & $\eta \sim 10^{-11}$ & $\eta \sim 10^{-8}$\\
\hline
\end{tabular}
\footnotetext{$^{a}$Efficiency is estimates as a ratio of the THz pulse energy to the total energy of optical
pulses absorbed in the crystal $\eta=E_{THz}/2E_{opt}$. The ultimate efficiency of the
method is given by the ratio of the THz radiation frequency to the frequency of the optical transitions
$29$ cm$^{-1}$/$14420$ cm$^{-1}$=$2\cdot10^{-3}$.}
\end{ruledtabular}
\end{table*}
\end{center}

At low temperatures as was already discussed above $55\%$ of optical intensity is absorbed at $L_{opt}=0.05$ cm.
Taking $A_{opt}=1$ cm$\times$ 0.1 cm we obtain from Eq.(\ref{eq:opt-energy}) for short pulses
$E_{opt}\approx 13.5$ mJ in one pulse, giving total required energy of $54$ mJ. For long pulses the estimate
gives $E_{opt}=67.5$ mJ in one pulse and $270$ mJ total energy for long pulses.

The parameters of THz pulses: energy, peak amplitude if focused to a spot of $300$ $\mu$m size, and
required optical pulses energy and duration at low temperatures are summurized in Table II.
\begin{center}
\begin{table}
\caption{\label{tab:TableII}Estimates of the THz pulse energy, peak THz field amplitude, required optical pulse parameters and efficiency of THz radiation
generation in ruby at low temperatures}
\begin{ruledtabular}
\begin{tabular}{cc}
\hline
ultrashort pulses (10 ps-1 ns)  & long pulses (1 ns-10 ns)\\
\hline
$\sigma_{cb}=0.5$ & $\sigma_{cb}=1$\\
$E_{THz}\sim 7.5\;\mu $J & $E_{THz}\sim 8\;\mu$J\\
${\cal{E}}_{THz}\sim 230\;$kV/cm & ${\cal{E}}_{THz}\sim 170\;$kV/cm\\
\hline
$\Omega_{0}=10^{10}\;s^{-1}$ & $\Omega_{0}\sim 5\cdot 10^{9}\;s^{-1}$\\
$I_{peak}=100\;$MW/cm$^{2}$ & $I_{peak}=30\;$MW/cm$^{2}$\\
$E_{opt}=54\;$mJ & $E_{opt}\sim 270\;$mJ\\
$\eta \sim 10^{-4}$ & $\eta \sim 3\cdot 10^{-5}$\\
\hline
\end{tabular}
\end{ruledtabular}
\end{table}
\end{center}

\section{DISCUSSION}

\subsection{Application to other solid materials}

It is important to note that the proposed technique can be applied to other
solid materials with suitable transitions in the THz range.
Similar schemes can be found in other ions as well, since for a typical
rare-earth or transition metal ion the
level structure is complex, having many crystal field and spin-orbit split
components, separated by tens to hundreds
of wave numbers. For example, this method is readily applicable to
Cr$^{3+}$:BeAl$_{2}$O$_{4}$ (alexandrite), having
R-lines splitting of 41 cm$^{-1}$ ($1.23$ THz) in a non-inversion Cr$^{3+}$
site \cite{alexandrite}, and R-lines
absorption/emission cross-section ten times larger than ruby
($3\cdot 10^{-19}$ cm$^{2}$ at room temperature).
It allows one to decrease peak optical intensities to induce maximal
THz coherence and results in higher convertion efficiency.

The technique can also be applied to rare-earth ion doped materials,
for example Pr$^{3+}$ doped hosts, such as LaF$_{3}$ \cite{Pr-LaF3},
Y$_{2}$SiO$_{5}$ \cite{Pr-YSO}, CaF$_{2}$ \cite{Pr-CaF2} with the ground and
excited state Stark splittings being in a broad range 17 - 100 cm$^{-1}$ and
higher, the same splittings are found in Nd$^{3+}$ doped crystals, such as
YAG \cite{Nd-YAG} and YVO$_{4}$ \cite{Nd-YVO}, and a number of others, for
example Tm:YAG \cite{Tm} and Er:YAG
\cite{Er-YAG} and Er:YLiF$_{4}$ \cite{Er-YLF}.

Also it is important to note that the technique can be applied to gases
as well, but the coupling to THz
radiation in this case occurs by a magnetic-dipole moment (macroscopic
polarization due to an electric-dipole moment is zero because of symmetry
reasons).
Typically the magnetic dipole moment is smaller than the electric dipole one,
so efficiency is smaller, but for higher densities and pressures this
approach can still produce THz radiation. The dipole moments for optical
transitions can be stronger than for ruby, so the coherence can be
larger. This possibility is an extension of the method proposed in
\cite{Yuri-Kolya1,Yuri-Kolya2}, although its analysis is beyond of the scope
of the present paper.

The next important remark is related to the case when the THz
transition of interest belongs to the ground state. Then there are a lot
of new opportunities to use more sophisticated methods of coherence
preparation.  In particular, the method of pulsed coherence production via
Stimulated Raman adiabatic passage (STIRAP) \cite{stirap} can be used to
control the duration of the THz pulse similarly to
\cite{Yuri-Kolya2}. Generally, all methods for coherence preparation via
femtosecond pulse shaping developed to improve the sensitivity of coherent Raman
scattering \cite{MaxCoh}, in particular, frequency chirping\cite{MaxCoh}
and fractional STIRAP~\cite{sautenkov03pra},
matched pulses~\cite{beadie05oc}, etc. can be applied.

\subsection{Comparison with current methods of short THz pulse generation}

It is interesting to compare the method of THz generation proposed in recent works \cite{Yuri-Kolya1,Yuri-Kolya2} and
in this paper with currently available ones. There are a variety of methods already considered successful for generation
of short THz pulses, so it is worth to compare the above estimates with the currently achieved parameters.

At the moment the most impressive results in generation of short THz pulses were achieved using electron beam
based sources, free-electron lasers (FEL) \cite{FEL1,FEL2} and synchrotrons \cite{synch1,synch2}. For FELs, typically
THz pulses of about 1 ps duration with 1-40 ns distance between micropulses are generated, grouped into a few-$\mu $s
trains. Energies of the micropulses are about 1-50 mJ (the parameters of FELBE laser, Rossendorf, Germany). Recently,
substantially higher powers of coherent broadband THz pulses, produced by synchrotron emission, were obtained
from the electron beamline \cite{synch1,synch2}. Hundreds of fs-short half-cycle THz pulses were generated with
energies up to 100 $\mu $J.

THz pulses with ns durations were achieved in THz semiconductor lasers (quantum cascade lasers, p-Ge and n-Si lasers
\cite{QC1}-\cite{QC5}), limited by the need of cryogenic cooling and in the case of p-Ge lasers by high electric and
magnetic fields required for laser operation. In p-Ge lasers \cite{QC2}-\cite{QC4} 20 ps pulse
durations were obtained in a mode-locked regime with peak power up to several Watts in a few-$\mu $s train of pulses.

The obvious advantages of the optical crystal based THz emitters discussed in
this work over FEL and synchrotron sources are compactness (typical optical solid-state materials are
significantly smaller), cost, and ease of handling.
Compared to THz QC lasers the crystals are much easier to grow and handle, they also can provide better
coupling between the
THz mode and the generating material due to a larger size of crystals.

\begin{figure*}
\center{
\includegraphics{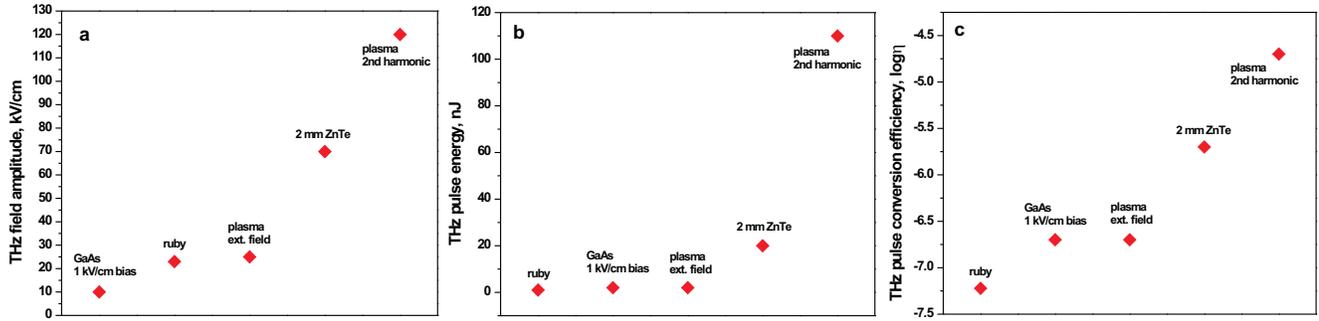}
\caption{\label{fig:comp-1} Predicted in \cite{amplif-laser-THz} performance of several amplifier-laser-based
THz sources and ruby at room temperature with optical pulses energy $\sim 10$ mJ: a) peak THz field amplitude; b)
THz pulse energy; c) conversion efficiency.}
}
\end{figure*}

Other methods of short THz pulse generation are based on interaction of different materials with ultrashort laser
pulses. The most popular ones, giving subpicosecond THz pulses, are the photocurrent method using the
Auston-switch technique \cite{Zhang} and the optical rectification technique \cite{Opt-Rectif}. Since the proposed
crystal-based THz sources are closer in their characteristics to these methods, we will make a more detailed comparison.
Recently THz sources based on amlifier-laser systems such as Ti:Sapphire, utilizing different methods of THz
radiation generation such as optical rectification in nonlinear crystals and laser-produced plasma and
photocurrent THz generation in semiconductor antennas, became popular.
The transition metal and rare-earth ion doped crystals proposed in this work as THz emitters are suitable for
amplifier-laser-based THz systems working
in a single-shot regime \cite{amplif-laser-THz}, since the typically long (hundreds $\mu $s - milliseconds) population decay time of excited
states of these ions limits the repetion rate. Crystals are also better in terms of saturation, able to withstand
high laser pulse fluence and thus very compatible with low-repetiotion rate table-top laser systems producing pulses
with energies in the 10-100 mJ range with $\sim 10$ Hz repetion rates. Fig.\ref{fig:comp-1} compares the predicted
in \cite{amplif-laser-THz} performance of different amplifier-laser-based THz sources and ruby at room temperature,
at optical pulses energy of $\sim 10$ mJ. It shows that ruby is expected to perform as good as the biased GaAs antenna
with 1 kV/cm dc bias field and the plasma THz source with an external bias field, slightly yielding them in
conversion efficiency.

\begin{figure*}
\center{
\includegraphics{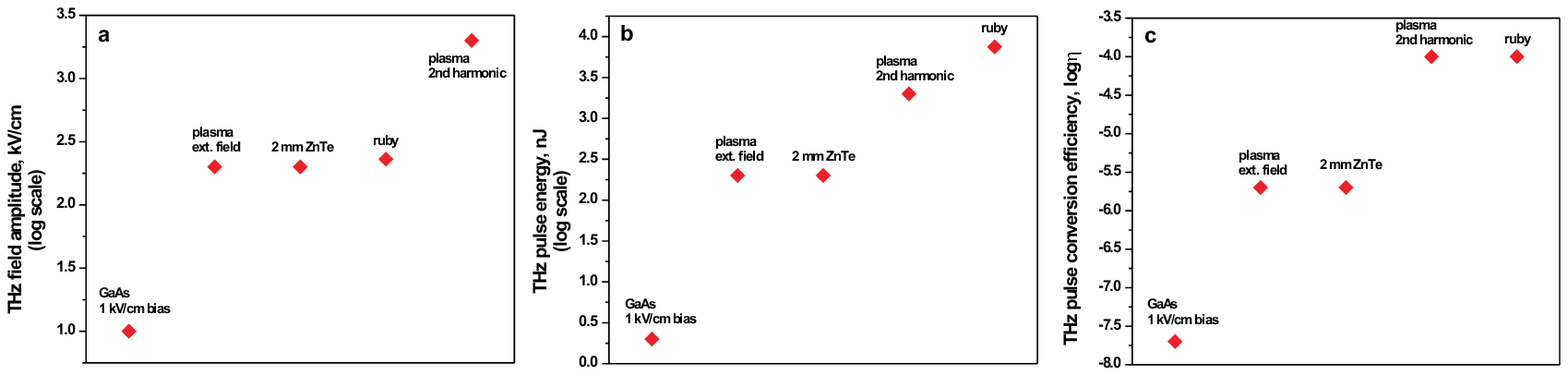}
\caption{\label{fig:comp-2} Predicted in \cite{amplif-laser-THz} performance of several amplifier-laser-based
THz sources at room temperature and ruby at liquid helium one with optical pulses energy $\sim 100$ mJ: a) peak THz field amplitude; b)
THz pulse energy; c) conversion efficiency.}
}
\end{figure*}

Fig.\ref{fig:comp-2} compares the predicted
in \cite{amplif-laser-THz} performance of different amplifier-laser-based THz sources at room temperature and ruby at
liquid helium one,
at optical pulses energy of $\sim 100$ mJ. The ruby source at liquid helium temperature can compete with the the plasma
source utilizing the fundumental and second harmonic optical fields in terms of THz pulse energy and conversion
efficiency, and yields it in peak THz field amplitude.

\subsection{Advantages of the technique}

New methods of THz generation proposed in \cite{Zibrov,Yuri-Kolya1,Yuri-Kolya2} for atomic and molecular gases and in the present work
for doped crystals potentially have very high
efficiency, they offer the possibility of THz pulses with controlled durations including femtosecond region and the
possibility of generation of pulses with high energy, already comparable with synchrotron or free-electron laser based
THz sources. Implementation of the method, therefore, will open exciting opportunities in many fields of THz applications
such as the time-resolved THz spectroscopy. Using intense short pulses of THz radiation generated by the proposed method
will significantly increase the performance of spectroscopic measurements, allowing maximal (among currently existing
methods) temporal resolution. Extremely high efficiency of the proposed technique opens a way for orders of magnitude
increase in sensitivity of the spectroscopic methods, THz tomography, non-destructive quality control, medical
diagnostics and biomaterial characterization.

Another advantage of the proposed scheme is a better control over the phase of the generated THz radiation, resulting
in lower (pulse duration $\times$ bandwidth) product compared to the non-coherent techniques. Resulting THz pulses
have narrower bandwidth, thus offering an opportunity of better resolution measurements.
Coherence of the generated THz pulses combined with their high energy opens the way to various non-linear phenomena in the
THz range, as well as to coherent phenomena similar to the ones observed with lasers in the visible range, such as
Rabi oscillations, coherent transients, etc.
The proposed methods have, therefore, high potential for investigation of nonlinear effects in chemical and biological
objects, in medical applications, which is a substantial, practically unexplored field of research.

\section{CONCLUSION}

In the present work we proposed a new technique for generation of short intense
THz pulses in coherently driven doped optical
crystals. The method is based on excitation of maximal THz coherence by a pair
of resonant optical pulses, resulting
in polarization build-up in the medium, which will radiate a THz pulse. As
an example, a well-known laser material
Cr$^{3+}$:Al$_{2}$O$_{3}$ (ruby) is considered and numerical estimates of
expected THz and required optical field parameters
are given. A number of other doped crystals, having transitions in the THz range, are suggested.
Comparison is made with the existing methods of short THz pulses
generation.

In summary, the optical crystal based THz sources proposed in this work can
potentially produce short high energy
(hundreds pJ - tens $\mu $J) THz pulses with durations ranging from 1 ps up to
several ns. The generated THz radiation
will be coherent, thus
offering the possibility of coherent interaction between the THz pulses and
probed media. This allows one to study
various types of nonlinear penomena. Obvious advantages of crystals are their compact size, ease of growth and handling,
robustness and low cost.

\section{ACKNOWLEDGEMENTS}

We gratefully acknowledge the support from the Air Force Office of Scientific
Research, the National Science Foundation, the Defense Advanced Research
Projects, the Office of Naval Research under
Award No. N00014-03-1-0385, and the Robert A. Welch Foundation (Grant
No. A1261).

\begin{thebibliography}{99}

\bibitem{THz1} {\it Terahertz optoelectronics}, edited by K. Sakai
(Springer, Berlin, 2005); B.Ferguson, and X.-C.Zhang, Nature Materials {\bf 1}, 26 (2002).

\bibitem{THz2} {\it Sensing with terahertz radiation}, edited by D.Mittleman
(Springer, New York, 2003).

\bibitem{THz3} N.G.Kalugin, in {\it Handbook of semiconductor nanostructures
and nanodevices}, edited by A.A.Balandin
and K.L.Wang (American Scientific Publishers, Los Angeles, 2005).

\bibitem{Zibrov} A. S. Zibrov, M. D. Lukin, L. Hollberg, and M. O. Scully,
Phys. Rev. A {\bf 65}, 051801(R) (2002).

\bibitem{Yuri-Kolya1} N.G.Kalugin, Y.Rostovtsev, M.O.Scully,
Proc. SPIE {\bf 6120}, 612002 (2006); N.G.Kalugin, Y.Rostovtsev, M.O.Scully,
quant-phys/0602142 (2006).

\bibitem{Yuri-Kolya2} N.G.Kalugin, Y.Rostovtsev, Opt. Lett.
{\bf 31}, 969 (2006).

\bibitem{lwi} 
O.~Kocharovskaya and Ya.I.~Khanin, Pis'ma Zn. Eksp. Teor. Fiz. {\bf 48}, 581
(1988), (Sov. Phys. JETP Lett. {\bf 48}, 630 (1988));
S.E.~Harris, Phys. Rev. Lett. {\bf 62}, 1033 (1989);
M.O.~Scully, S.Y.~Zhu, and A.~Gavrielides,
Phys. Rev. Lett. {\bf 62}, 2813 (1989).

\bibitem{EIT} E.~Arimondo in {\em Progress in Optics XXXV},
ed. by E.~Wolf (Elsevier, Amsterdam, 1996), p.257;
S.E.Harris, Phys. Today {\bf 50}, 36 (1997).

\bibitem{LWI-exp} 
A.S.~Zibrov,  M.D.Lukin, D.E.Nikonov, L.Hollberg, M.O.Scully, V.L.Velichansky, H.G.Robinson, Phys. Rev. Lett. { \bf 75},
1499 (1995);
G.G.~Padmabandu, G.R.Welch, I.N.Shubin, E.S.Fry, D.E.Nikonov, M.D.Lukin, M.O.Scully, Phys. Rev. Lett. {\bf 76},
2053 (1996).

\bibitem{EIT-exp} G.Alzetta, A.Gozzini, L.Moi, G.Orriols, Nuovo Cimento B {\bf 36}, 5 (1976);
H.R.Gray, R.M.Whitley, C.R.Stroud, Jr., Opt. Lett. {\bf 3}, 218 (1978); K.J.Boller, A.Imamoglu, S.E.Harris, Phys. Rev.
Lett. {\bf 66}, 2593 (1991); J.E.Field, K.H.Hahn, S.E.Harris, Phys. Rev. Lett. {\bf 67}, 3062 (1991).

\bibitem{Harris-opt-param} S.E.Harris, M.Jain, Opt. Lett. {\bf 22}, 636 (1997).

\bibitem{Boyd-Scully} R.W.Boyd, M.O.Scully, Appl. Phys. Lett. {\bf 77}, 3559 (2000).

\bibitem{Harris} M.Jain, H.Xia, G.Y.Yin, A.J.Merriam, S.E.Harris, Phys. Rev. Lett. {\bf 77}, 4326 (1996).

\bibitem{kocharovskaya99prl} O. Kocharovskaya, R. Kolesov, and Yu. Rostovtsev,
Phys. Rev. Lett. {\bf 82}, 3593-3596 (1999)


\bibitem{sautenkov03pra} V. A. Sautenkov, C. Y. Ye, Y. V. Rostovtsev,
  G. R. Welch, and M. O. Scully, Phys. Rev. A {\bf 70}, 033406 (2004).

\bibitem{MaxCoh} M.O.Scully, G.W.Kattawar, P.R.Lucht, T.Opatrny, H.Pilloff, A.Rebane, A.V.Sokolov, M.S.Zubairy,
Proc. Natl. Acad. Sci. U.S.A. {\bf 9}, 10994 (2002).

\bibitem{kuznetsova02pra}
E. Kuznetsova, O. Kocharovskaya, P. Hemmer, and M. O. Scully,
Phys. Rev. A {\bf 66}, 063802 (2002).

\bibitem{kuznetsova04pra}
 E. Kuznetsova, R. Kolesov, and O. Kocharovskaya,
 Phys. Rev. A {\bf 70}, 043801 (2004).

\bibitem{kolesov05pra}
R.Kolesov, E. Kuznetsova, and O. Kocharovskaya,
 Phys. Rev. A {\bf 71}, 043815 (2005).


\bibitem{lwi06nature} M.D. Frogley, J.F. Dynes, M. Beck, J. Faist,
  C.C. Phillips, Nature Mat. {\bf 5} 175 (2006).


\bibitem{beadie05oc} G. Beadie, Z.E. Sariyanni, Y.V. Rostovtsev,
T. Opatrny, J. Reintjes, M.O. Scully, Opt. Commun.  {\bf 244}, 423 (2005).

\bibitem{stirap} K. Bergmann, H. Theuer, and B. W. Shore,
 Rev. Mod. Phys. {\bf 70}, 1003 (1998).



\bibitem{Dima-Vitaly} D.S.Pestov, A.A.Belyanin, V.V.Kocharovsky, Vl.V.Kocharovsky, M.O.Scully, J. Mod. Opt. {\bf 51}, 2523 (2004).

\bibitem{sel-rules} M.O.Schweika-Kresimon, J.Gutschank, D.Suter, Phys. Rev. A
  {\bf 66}, 043816 (2002).

\bibitem{FID} R.G.Brewer, R.L.Shoemaker, Phys. Rev. A {\bf 6}, 2001 (1972).

\bibitem{ruby} D.F.Nelson, and M.D.Sturge, Phys. Rev. {\bf 137}, A1117 (1965).

\bibitem{quant-counter} H.Lengfellner, K.F.Renk, IEEE J. Quant. Elect. {\bf
  13}, 421 (1977).

\bibitem{ruby-phonons} A.A.Kaplyanskii and S.A.Basun, in {\it Nonequilibrium Phonons in Nonmetallic Crystals}, edited
by W.Eisenmenger and A.A.Kaplyanskii (North-Holland, Amsterdam, 1986), p.373.

\bibitem{FIR-laser} N.M.Lawandy, IEEE J. Quant. Elect. {\bf 15}, 401 (1979).

\bibitem{FIR-mixing} D.W.Faries, P.L.Richards, Y.R.Shen, K.H.Yang, Phys. Rev. A {\bf 3}, 3 (1971).

\bibitem{ZnTe} T.Yajima, K.Inoue, Phys. Lett. A {\bf 26}, 281 (1968).

\bibitem{FIR-room-temp} B.Halperin, J.A.Koningstein, D.Nicollin, Chem. Phys. Lett. {\bf 68}, 58 (1979).

\bibitem{ruby-damage-threshold} S.A.Belozerov, G.M.Zverev, V.S.Naumov, V.A.Pashkov, Soviet Physics JETP {\bf 35}, 158 (1972).

\bibitem{ruby-damage-threshold1} G.M.Zverev, T.N.Mikhailova, V.A.Pashkov, N.M.Solov'eva, Soviet Physics JETP {\bf 26}, 1053 (1968).

\bibitem{FIR-absorp} N.Retzer, H.Lengfellner, and K.F.Renk, Phys. Lett. {\bf 96A}, 487 (1983).

\bibitem{ruby-R-line} D.E.McCumber, M.D.Sturge, J. Appl. Phys. {\bf 34}, 1682 (1963).

\bibitem{phonons} H.Lengfellner, J.Hummel, H.Netter, K.F.Rank, Opt. Lett. {\bf 8}, 220 (1983).

\bibitem{alexandrite} J.C.Walling, O.G.Peterson, H.P.Jenssen, R.C.Morris, E.W.O'Dell, IEEE J. Quant. Electr. {\bf 16}, 1302 (1980).

\bibitem{Pr-LaF3} R.M.Shelby, R.M.Macfarlane, C.S.Yannoni, Phys. Rev. B {\bf 21}, 5004 (1980).

\bibitem{Pr-YSO} R.W.Equall, R.L.Cone, R.M.Macfarlane, Phys. Rev. B {\bf 52}, 3963 (1995).

\bibitem{Pr-CaF2} R.M.Macfarlane, D.P.Burum, R.M.Shelby, Phys. Rev. B {\bf 29}, 2390 (1984).

\bibitem{Nd-YAG} G.W.Burdick, C.K.Jayasankar, F.S.Richardson, M.F.Reid, Phys. Rev. B {\bf 50}, 16309 (1994).

\bibitem{Nd-YVO} D.K.Sandar, R.M.Yow, Opt. Mater. {\bf 14}, 5 (2000).

\bibitem{Tm} R.M.Macfarlane, J. of Lumin. {\bf 85}, 181 (2000).

\bibitem{Er-YAG} J.B.Gruber, J.R.Quagliano, M.F.Reid, F.S.Richardson, M.E.Hills, M.D.Seltzer, S.B.Stevens, C.A.Morrison, T.H.Allik, Phys. Rev. B {\bf 48}, 15561 (1993).

\bibitem{Er-YLF} R.A.Macfarlane, J. Opt. Soc. Am. B {\bf 8}, 2009 (1991).

\bibitem{FEL1} W.Chin, et al., J. Phys. Chem. {\bf 122}, 054317 (2005).

\bibitem{FEL2} X.G.Peralta, et al., Appl. Phys. Lett. {\bf 81}, 1627 (2002).

\bibitem{synch1} C.R.Neil, et al., Nucl. Instr. Meth. Phys. Res. A {\bf 507}, 537 (2003).

\bibitem{synch2} G.L.Carr, et al., J. Biol. Phys. {\bf 29}, 319 (2003).

\bibitem{QC1} R.Koehler, A.Tredicucci, F.Bertram, H.E.Beere, E.H.Linfield, A.G.Davies, D.A.Ritchie, R.C.Iotti, R.F.Rossi, Nature {\bf 417}, 156 (2002).

\bibitem{QC2} A.A.Andronov, I.V.Zverev, V.A.Kozlov, Yu.N.Nozdrin, S.A.Pavlov, V.N.Shastin,
JETP Lett. {\bf 40}, 804 (1984); E.Gornik, A.A.Andronov (ed), Opt. Quant. Electron. {\bf 23}, S111 (1991).

\bibitem{QC3} V.I.Gavrilenko, N.G.Kalugin, Z.E.Krasilnik, V.V.Nikonorov, A.V.Galyagin, P.N.Tsereteli, Semicond. Sci. Technol. {\bf 7}, B649 (1992).

\bibitem{QC4} A.V.Muravjov, R.C.Strijbos, C.J.Fredricksen, H.Weidner, W.Trimble, S.G.Pavlov, V.N.Shastin, R.E.Peale, Appl. Phys. Lett. {\bf 73}, 3037 (1998).

\bibitem{QC5} S.G.Pavlov, R.K.Zhukavin, E.E.Orlova, V.N.Shastin, A.V.Kirsanov, H.W.Hubers, K.Auen, H.Riemann, Phys. Rev. Lett. {\bf 84}, 5220 (2000).

\bibitem{Zhang} X.-C.Zhang, B.B.Hu, J.T.Darrow, D.H.Auston, Appl. Phys. Lett. {\bf 56}, 1011 (1990); D.H.Auston, K.P.Cheung, P.R.Smith, Appl. Phys. Lett. {\bf 45}, 284 (1984).

\bibitem{Opt-Rectif} K.H.Yang, P.L.Richards, Y.R.Shen, Appl. Phys. Lett. {\bf 19}, 320, (1971); A.Rice, X.F.Ma, X.-C.Zhang, D.Bliss, J.Larkin, M.Alexander, Appl. Phys. Lett. {\bf 64}, 1324 (1994).

\bibitem{amplif-laser-THz} T.L\"offer, M.Kreb, M.Thomson, T.Hahn, N.Hasegawa, H.G.Roskos, Semicond. Sci. Technol. {\bf 20}, S134 (2005).

\end{thebibliography}
\end{document}